\documentclass[preprint,aps,showpacs,nofootinbib,preprintnumbers,amsmath,amssymb]{revtex4-1}
\usepackage{epsfig}

\usepackage{dcolumn}
\usepackage{bm}
\usepackage[usenames ,dvipsnames]{xcolor}
\usepackage{slashed}
\usepackage{graphicx,color}
\usepackage{caption}
\usepackage{subcaption}
\begin{document}
\title{ Singly Cabibbo suppressed decays of $\Lambda_{c}^+$ with  SU(3) flavor symmetry}

\author{Chao-Qiang Geng$^{1,2,3}$, Chia-Wei Liu$^{2}$ and  Tien-Hsueh Tsai$^{2}$}
\affiliation{
$^{1}$Chongqing University of Posts \& Telecommunications, Chongqing 400065\\
$^{2}$Department of Physics, National Tsing Hua University, Hsinchu 300\\
$^{3}$Physics Division, National Center for Theoretical Sciences, Hsinchu 300
}\date{\today}

\begin{abstract}
We analyze the weak processes of anti-triplet charmed baryons decaying to octet baryons and mesons with the SU(3) flavor symmetry and topological quark diagram scheme. We study the decay branching ratios without neglecting the contributions from ${\cal O}(\overline{15})$
for the first time in the SU(3) flavor symmetry approach.
The fitting results for the  Cabibbo allowed and suppressed   decays of $\Lambda_{c}^+$ are  all consistent with the experimental data.
We  predict all singly Cabibbo  suppressed  decays.  In particular,
we  find that ${\cal B}(\Lambda_c^+\to p \pi^0)=(1.3\pm0.7)\times 10^{-4}$, which is slightly below the current experimental upper
limit of $2.7\times 10^{-4}$ and can be tested by the ongoing experiment at BESIII as well as the future one at Belle-II.
\end{abstract}

\maketitle
Recently, the study of the charmed baryons has been receiving increasing attention both theoretically and experimentally.
The main reason for this is the recent measurement of the absolute branching fraction of the golden channel $\Lambda_c^+ \to p K^-\pi^+$
by the Belle Collaboration~\cite{Belle}. This mode and many other $\Lambda_c^+$ ones have also been observed
by the BESIII Collaboration~\cite{Ablikim:2015flg,Ablikim:2015prg,Ablikim:2016tze,Ablikim:2016mcr,Ablikim:2017ors,
Ablikim:2016vqd,Ablikim:2017iqd,Ablikim:2018jfs,Ablikim:2018bir} with  using $\Lambda_c^+\bar{\Lambda}_c^-$ pairs produced by $e^+ e^-$ collisions
at a center-of-mass energy of $\sqrt{s}= 4.6$ GeV,  which provides a uniquely clean background to study charmed baryons.
Consequently, the Particle Data Group (PDG)~\cite{pdg} has given  a new average of ${\cal B}(\Lambda_c^+ \to p K^-\pi^+)=(6.23\pm0.33)\%$.
The precision measurement on this mode  is very important as it can be used to determine the absolute branching fractions of
other $\Lambda_c^+$ decays~\cite{pdg} as well as  processes involving $\Lambda_c^+$, such as the extractions of the CKM element from
$\Lambda_b\to \Lambda_c^+\mu^-\bar{\nu}_\mu$~\cite{Aaij:2015bfa,Hsiao:2018zqd}.
It is clear that a new era of physics for charmed baryons has begun.
 For a review on the theoretical progress of charmed baryons, please see Ref.~\cite{Cheng:2015iom}.

On the other hand,
the singly Cabibbo suppressed decays of $\Lambda_{c}^+\to p \eta$ and $\Lambda_c^+ \to p \pi^0$ have been recently
investigated by BESIII ~\cite{Ablikim:2017ors}. The branching fraction of the former mode
has been measured for the first time with
${\cal B}(\Lambda_{c}^+\to p \eta)=(1.24\pm0.28\pm 0.10)\times10^{-3}$,
whereas that of the later one  has also been searched with no significant  signal  observed,
resulting in  an upper limit of  ${\cal B}(\Lambda_c^+ \to p \pi^0) < 2.7\times 10^{-4}$ at the 90\% confidence level.
These two decays have been extensively studied in the literature based on various dynamical
models~\cite{Uppal:1994pt,Chen:2002jr,Cheng:2018hwl}
 as well as the flavor $SU(3)_F$ symmetry~\cite{Sharma:1996sc,Lu:2016ogy,zero,first,second,third}.
In particular, Cheng, Kang and Xu (CKX)~\cite{Cheng:2018hwl} have performed a dynamical calculation based on current algebra to examine
 the decay of $\Lambda_c^+ \to p \pi^0$ and found that its branching fraction is $0.8\times 10^{-4}$, which is consistent with the current
 experimental upper limit. However, those with $SU(3)_F$ have given an inconsistent  larger value,
 e.g., $(5.7\pm1.5)\times 10^{-4}$ in Ref.~\cite{second}.

It is known that it is difficult to make reliable predictions on the charmed baryon decay rates
due to  the lack of theoretical understanding of underlined dynamics for the charmed baryon structure.
 Since the Cabbibo allowed decays of $\Lambda_c^+ \to \Sigma^0\pi^+$ and $\Lambda_c^+ \to \Sigma^+\pi^0$ do not receive any factorizable contributions, the nonzero experimental observed values of their branching fractions
imply that the factorization approach is not working in charmed baryon decays.
Without the use of a dynamical model, it is clear that
the most reliable way to analyze charmed baryon processes is to impose 
$SU(3)_F$ 
 ~\cite{Sharma:1996sc,Lu:2016ogy,first,zero,second,third,fourth,Savage:1989qr,Savage:1991wu}.
 In fact, it has been demonstrated~\cite{first,zero,second,third} that all the existing data of the Cabbibo favored and suppressed charmed baryon decays
  except ${\cal B}(\Lambda_c^+ \to p \pi^0)$ can be fitted  well.
  In these calculations under $SU(3)_F$, the contributions to the decays from the sextet ${\bf 6}$ are assumed to be the dominant ones, whereas those from
  ${\bf \overline{15}}$ are neglected, by taking into account of  the enhancements of the QCD running Wilson coefficients associated with the
  sextet ${\bf 6}$ part~\cite{old1,old2,Abbott:1979fw,Zeppenfeld:1980ex}
  and the vanishing baryonic transition matrix elements from the nonfactorizable contributions
   with  ${\bf \overline{15}}$~\cite{Cheng:2018hwl}.
 However, it is interesting to ask what the contributions to the decay rates from the factorizable parts of  ${\bf \overline{15}}$ are.
In this note, we will try to answer this question. Specifically, we examine all possible contributions to the charmed baryon decays under $SU(3)_F$
without neglecting those from ${\bf \overline{15}}$.
We examine the singly Cabibbo suppressed  $\Lambda_{c}^+$ decays to check if our results are consistent with the data,
in particular, the $\Lambda_c^+ \to p \pi^0$ channel.

There are two  approaches  to write down the irreducible decay amplitude through $SU(3)_F$. One
 is to generalize the Wigner Eckart theorem~\cite{deSwart:1963pdg} by writing  the decay amplitude
 to be invariant and singlet under  $SU(3)_F$.
 The other is to use  topological quark diagrams,
 where the decay amplitude is represented by all possible diagrams  connected by
quark lines which satisfy  $SU(3)_F$.
  Both  two have their own advantages.
For the former, one is able to compare the contributions from different representations of operators.
In this case, it is also possible to include the  $SU(3)_F$  breaking effect  by introducing
the strange quark mass~\cite{third,Savage:1991wu}.
On the other hand, the irreducible amplitude in the later approach is more intuitive and gives an insight on dynamics~\cite{Chau:1995gk}.
In particular, it could shed light for us on distinguishing the  nonfactorizable and factorizable contributions in the processes.
It is expected that these two approaches should give the same results under $SU(3)_F$.
The close connections between the two have been recently examined in Ref.~\cite{He:2018joe}.


To study the two-body anti-triplet of the lowest-lying charmed baryon decays of ${\bf B}_c\to {\bf B}_n M$,
where ${\bf B}_{c} = (\Xi_c^0,-\Xi_c^+,\Lambda_c^+)$
and  ${\bf B}_n$ and $M$ are the baryon and pseudoscalar octet states, given by
\begin{eqnarray}
 {\bf B}_n&=&\left(\begin{array}{ccc}
\frac{1}{\sqrt{6}}\Lambda+\frac{1}{\sqrt{2}}\Sigma^0 & \Sigma^+ & p\\
 \Sigma^- &\frac{1}{\sqrt{6}}\Lambda -\frac{1}{\sqrt{2}}\Sigma^0  & n\\
 \Xi^- & \Xi^0 &-\sqrt{\frac{2}{3}}\Lambda
\end{array}\right)\,,
\end{eqnarray}
\begin{eqnarray}
M&=&\left(\begin{array}{ccc}
\frac{1}{\sqrt{6}}\eta+ \frac{1}{\sqrt{2}}\pi^0  & \pi^+ & K^+\\
 \pi^- &\frac{1}{\sqrt{6}}\eta - \frac{1}{\sqrt{2}}\pi^0 &  K^0\\
 K^- & \bar K^0& -\sqrt{\frac{2}{3}}\eta
\end{array}\right)\,.
\end{eqnarray}
Here, we have assumed that the physical state of $\eta$ is solely made of $\eta_8$ due to the small mixing between the weak eigenstates of 
$\eta_0$ and $\eta_8$~\cite{pdg} to reduce our fitting parameters.

We start with the effective Hamiltonian responsible for the tree-level $c\to  s u\bar d$,
$c\to u q\bar q$ and $c\to  du\bar s$ transitions,
 given by~\cite{Buras:1998raa}
\begin{eqnarray}\label{Heff}
{\cal H}_{eff}&=&\sum_{i=+,-}\frac{G_F}{\sqrt 2}c_i
\left(V_{cs}V_{ud}O^{ds}_i+V_{cd}V_{ud} O^{qq}_i+V_{cd}V_{us}O^{sd}_i\right),
\end{eqnarray}
with
\begin{eqnarray}
\label{O12}
O_\pm^{q_2q_1}&=&{1\over 2}\left[(\bar u q_1)(\bar{q}_2c)\pm (\bar{q}_2 q_1)(\bar u c)\right]\,,
\end{eqnarray}
where $O_\pm^{q_2q_1}$ and $O_\pm^{qq}\equiv O_\pm^{dd}-O_\pm^{ss}$ are the four-quark operators,
$(\bar q_1 q_2)\equiv\bar q_1\gamma_\mu(1-\gamma_5)q_2$,
$G_F$  is the Fermi constant, $V_{ij}$ are the CKM matrix elements, and
$(c_+\,,c_-)  = (0.76\,,1.78)$, corresponding to the scale-dependent Wilson coefficients with the QCD corrections.
By using $(V_{cs}V_{ud},V_{cd}V_{ud},V_{cd}V_{us})\simeq (1,-s_c,-s_c^2)$ in Eq.~(\ref{Heff}) with $s_c\equiv \sin\theta_c=0.2248$~\cite{pdg}
representing  the well-known Cabbibo angle $\theta_c$, the decays associated with
$O_\pm^{ds}$, $O_\pm^{qq}$ and $O_\pm^{sd}$ are the so-called Cabibbo-allowed, singly Cabibbo-suppressed
and doubly Cabibbo-suppressed processes, respectively.

Under $SU(3)_F$, the operators in Eq.~(\ref{O12}) correspond to $(\bar q^i q_k)( \bar q^jc)$ with $q_i=(u,d,s)$ as the triplet of ${\bf 3}$,
which can be decomposed as the irreducible forms of
${(\bf \bar 3\times 3\times \bar 3})c=({\bf \bar 3+\bar 3'+6+\overline{15}})c$ with $c$ as a flavor singlet.
As a result, $(O_-,O_+)$ fall into the irreducible presentations of $({\cal O}_{6},{\cal O}_{\overline{15}})$~\cite{Savage:1989qr}.
In analogy to octet baryons and mesons, we can  write down the operators related to ${\cal O}_{6}$ and ${\cal O}_{\overline{15}}$
in tensor forms, given by
\begin{eqnarray}
\label{H6-15}
\left(H(6)_{ij}\right)&=&\left(\begin{array}{ccc}
0& 0 & 0\\
0 & 2 & -2s_c\\
0 & -2s_c& 2s_c^2
\end{array}\right)\,,\nonumber\\
\left(H(\overline{15})^{ij}_k\right)&=&
\left(\begin{array}{ccc}
\left(\begin{array}{ccc}
0&0&0\\
0&0&0\\
0&0&0
\end{array}\right),
\left(\begin{array}{ccc}
0&s_c&1\\
s_c&0&0\\
1&0&0
\end{array}\right),
\left(\begin{array}{ccc}
0&-s_c^2&-s_c\\
-s_c^2&0&0\\
-s_c&0&0
\end{array}\right)
\end{array}\right),
\end{eqnarray}
with $(i,j,k)$=1,2 and 3,
where  $H(\overline{15})$ is traceless and symmetric in upper indies, while $H(6)_{ij}$ is symmetric in lower indies. 
One can also write the matrix elements of $H(6)_{ij}$ and $H(\overline{15})^{ij}_k$ in Eq.~(\ref{H6-15}) as a single one, given by
\begin{equation}\label{rea}
H^{ij}_k=\frac{1}{2}\left(H(\overline{15})^{ij}_k+\frac{1}{2}\epsilon^{ijl}H(6)_{kl}\right)\,.
\end{equation}

Now, we can write down the SU(3) irreducible amplitude for ${\bf B}_c \to {\bf B}_n M$ as~\cite{first,Savage:1989qr}
\begin{eqnarray}
\label{Amp}
  {\cal A}({\bf B}_c \to {\bf B}_n M) &=& \langle {\bf B}_n M | {\cal H}_{eff}|{\bf B}_c \rangle
  \equiv \frac{G_F}{\sqrt{2}}\left(T_{{\cal O}_6}+T_{{\cal O}_{\overline{15}}}\right)\,,
\end{eqnarray}
where
\begin{eqnarray}\label{Tamp}
T_{{\cal O}_6}&=&
{a_1 H_{ij}(6)({\bf B}'_c)^{ik}({\bf B}_n)_k^l (M)_l^j+
a_2 H_{ij}(6)({\bf B}'_c)^{ik}(M)_k^l ({\bf B}_n)_l^j+
a_3 H_{ij}(6)({\bf B}_n)_k^i (M)_l^j ({\bf B}'_c)^{kl}}\nonumber\\
T_{{\cal O}_{\overline{15}}}&=&
a_4H_{k}^{li}(\overline{15})({\bf B}_{c})_j (M)_i^j ({\bf B}_n)_l^k
+a_5({\bf B}_n)^i_j (M)^l_i H(\overline{15})^{jk}_l ({\bf B}_{c})_k\nonumber\\
&&
+a_6({\bf B}_n)^k_l (M)^i_j H(\overline{15})^{jl}_i ({\bf B}_{c})_k
+a_7({\bf B}_n)^l_i (M)^i_j H(\overline{15})^{jk}_l ({\bf B}_{c})_k\,,
\end{eqnarray}
with $({\bf B}'_c)^{jk}\equiv({\bf B}_c)_i\epsilon^{ijk}$. Here,  the Wilson coefficients have been  absorbed in the parameters $a_i$.

\begin{figure}[t]
	\centering
	\vspace{-8em}
	\hspace{1.em}
	\begin{subfigure}{0.35\textwidth} 
		\includegraphics[width=\textwidth]
	{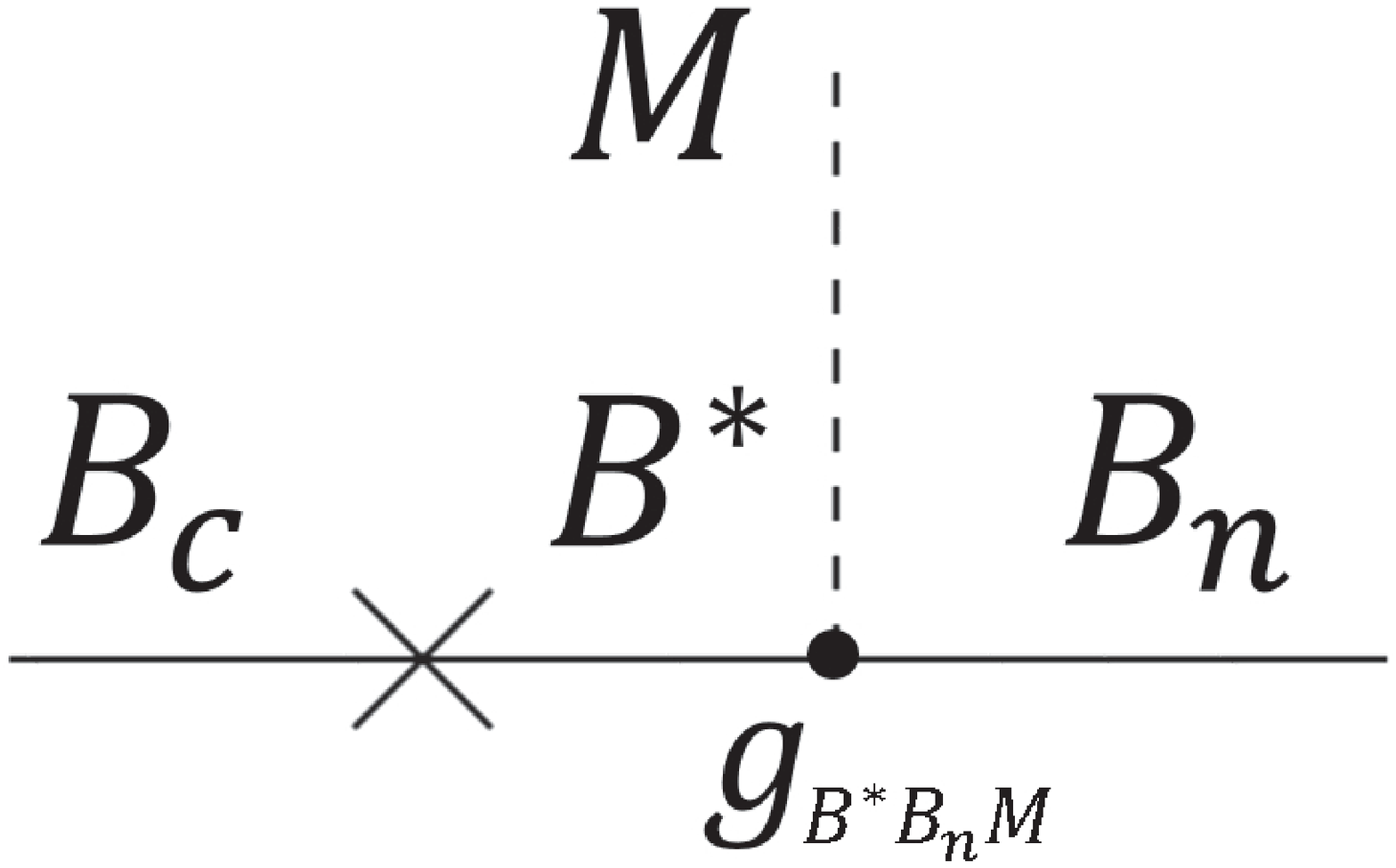}
	\end{subfigure}
	\vspace{-7em}
	\hspace{-1.em} 
	\begin{subfigure}{0.35\textwidth} 
		\includegraphics[width=\textwidth]
		{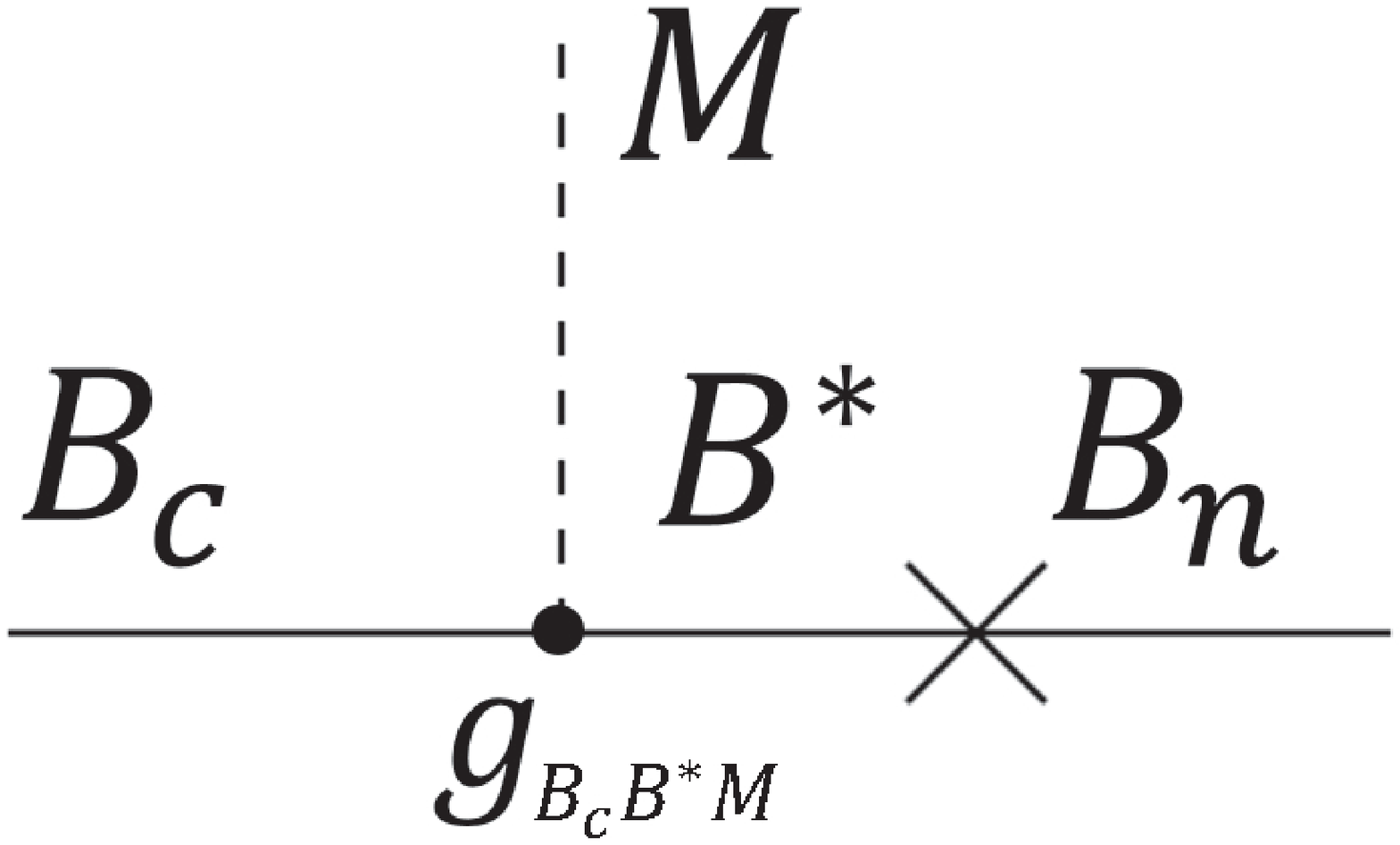}
	\end{subfigure}
	\caption{Pole diagrams of the nonfactorizable amplitude for ${\bf B}_c \to {\bf B}_n M$.}
	\label{FigB}
\end{figure}

\begin{figure}[!b]
	\centering
	\hspace{-2.5em}
		\begin{subfigure}{0.30\textwidth} 
		\includegraphics[width=\textwidth]
{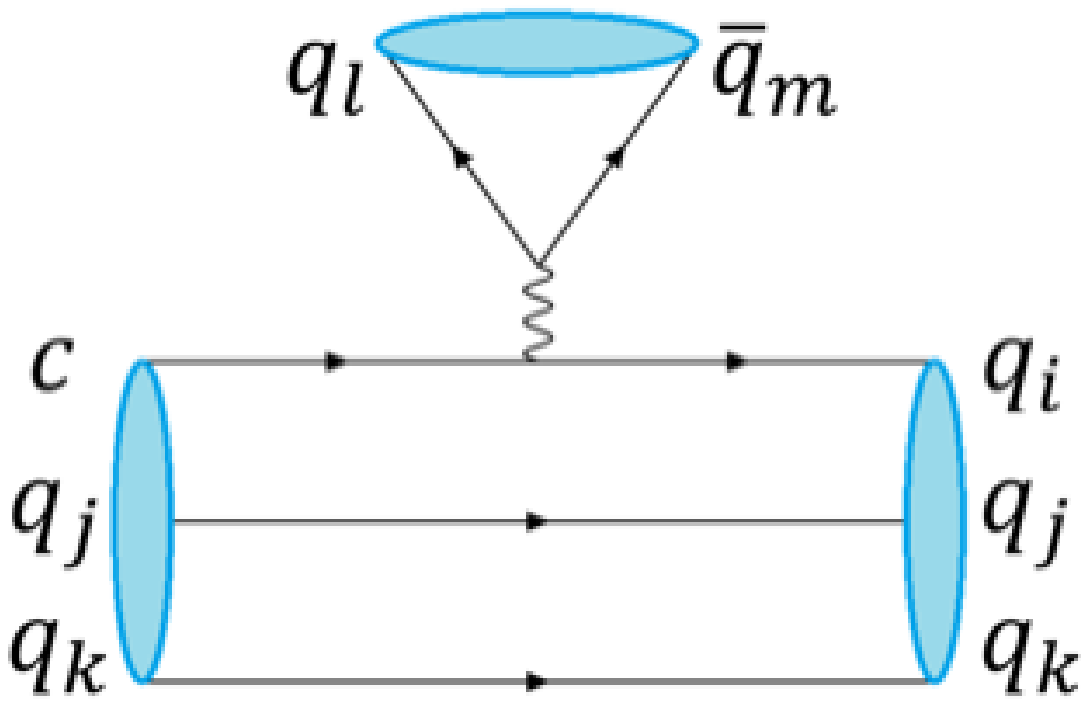}
		\caption{Color allowed diagram} 
	\end{subfigure}
\hspace{4.5em}
	\begin{subfigure}{0.33\textwidth} 
		\includegraphics[width=\textwidth]
		{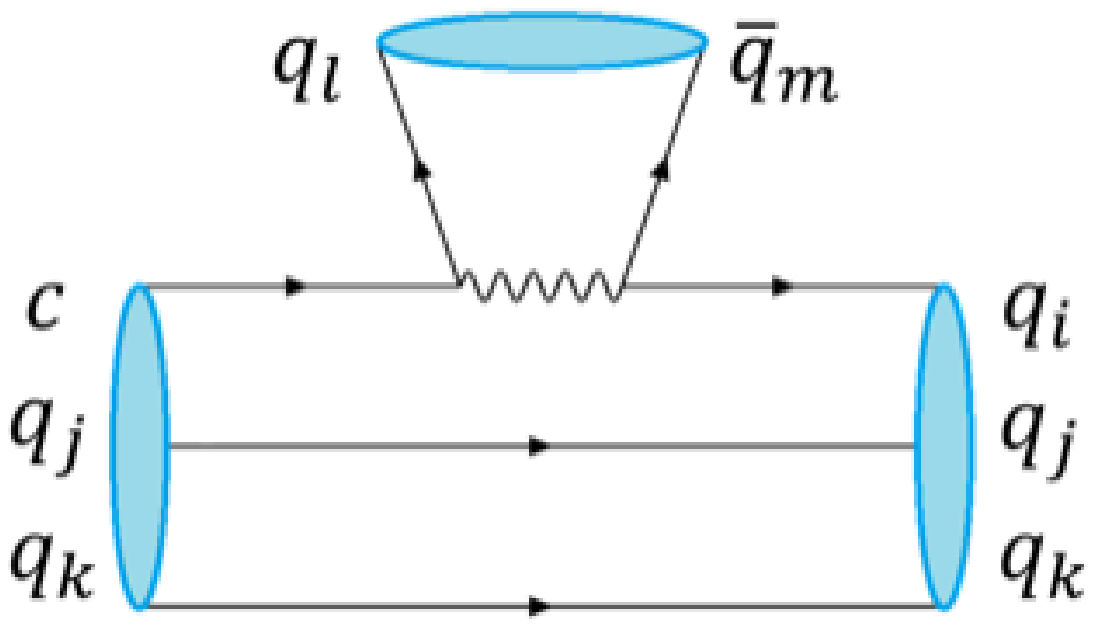}
		\caption{Color suppressed diagram} 
	\end{subfigure}
	\caption{Topological diagram for color allowed and suppressed processes.}
	\label{Fig1}
\end{figure}

In order to reduce the fitting parameters for the processes based on
 the amplitudes in Eqs.~(\ref{Amp}) and (\ref{Tamp}),
 as mentioned early,  the contributions related to ${\cal O}(\overline{15})$ in Eq.~(\ref{Amp}) have been neglected due to the fact that $c_-/c_+\approx 2.5$
and the vanishing contributions of ${\cal O}(\overline{15})$ from the nonfactorizable part to the amplitude.
To see the later reason, we write the amplitude of ${\bf B}_c\to {\bf B}_nM$ due to ${\cal O}(\overline{15})$  in terms of the matrix element
\begin{eqnarray}
\label{AO15}
{\cal A}({\cal O}(\overline{15})) &=& \langle {\bf B}_nM| {\cal O}(\overline{15})| {\bf B}_c\rangle
={1\over 2} \langle {\bf B}_nM| (\bar u q_1)(\bar{q}_2c)+ (\bar{q}_2 q_1)(\bar u c)| {\bf B}_c\rangle\,.
\end{eqnarray}
Since the operator ${\cal O}(\overline{15})\sim (\bar u q_1)(\bar{q}_2c)+ (\bar{q}_2 q_1)(\bar u c)$ is symmetric in color indices,
whereas  the baryon states ${\bf B}_{i}$ are  antisymmetric, one  easily arrives  that
$\langle {\bf B}_i|{\cal O}(\overline{15})|{\bf B}_j\rangle=0$.
From the calculations of the nonfactorizable (NF) contributions in terms of the baryon poles (${\bf B^*}$), 
one has that 
  ${\cal A}({\cal O}_{NF}(\overline{15}))$ is related to the combination of
   $g_{{\bf B_cB^*}M}\langle {\bf B}^*|{\cal O}(\overline{15})|{\bf B}_c\rangle$ and $g_{{\bf B^*B_n}M}\langle {\bf B}_n|{\cal O}(\overline{15})|{\bf B}^*\rangle$
   as illustrated in Fig.~\ref{FigB}~\cite{Cheng:1991sn}, indicating that  ${\cal O}(\overline{15})$ does not contribute  the nonfactorizable amplitude~\cite{Cheng:2018hwl}.
As a result, the amplitude in Eq.~(\ref{AO15}) only contains the factorizable (F) contributions
in the decays of ${\bf B}_c\to {\bf B}_nM$, and can be factorized as
\begin{eqnarray}
\label{AO15F}
{\cal A}({\cal O}(\overline{15})) &=& {\cal A}_F({\cal O}(\overline{15}))
\nonumber\\&=&
{1\over 2} \langle M| (\bar u q_1)|0\rangle \langle{\bf B}_n|(\bar{q}_2c)| {\bf B}_c\rangle
+
{1\over 2} \langle M|(\bar{q}_2 q_1)|0\rangle \langle{\bf B}_n| (\bar u c)| {\bf B}_c\rangle
\,.
\end{eqnarray}
To evaluate ${\cal A}_F({\cal O}(\overline{15}))$, we need the help of topological quark diagrams.
In other words, we have to find out the terms in $T({\cal O}_{\overline{15}})$ of Eq.~(\ref{Tamp}), which can be
 factorizable.
In  Figs.~\ref{Fig1}a and \ref{Fig1}b, we illustrate the factorizable contributions
for the color allowed and  suppressed processes in the topological diagram approach,\footnote{It is clear that we have ignored the soft gluon
interactions whenever the factorization problem is discussed.} respectively.
Note that  the quark indices represent the light quark lines of hadrons or operators with $q_i=(u,d,s)$.
From Fig.~\ref{Fig1}, we obtain that
\begin{equation}\label{TOP}
{\cal A}_F({\bf B}_c \to {\bf B}_n M)=T({\bf B}'_n)_{ijk}({\bf B}'_c)^{jk}H^{li}_mM^m_l + C ({\bf B}_n')_{ijk}({\bf B}_c')^{jk}H^{il}_mM^m_l
\end{equation}
where $({\bf B}'_n)_{ijk}\equiv ({\bf B}_n)^n_i\epsilon_{njk}$, $({\bf B}'_c)^{jk}\equiv({\bf B}_c)_m\epsilon^{mjk}$,
and
 $T(C)$ represents  the  color allowed  (suppressed) amplitude.
By using Eq.~\eqref{rea} and  the tensor identity $\epsilon_{njk}\epsilon^{mjk}=2\delta^m_n$, we find that
\begin{eqnarray}\label{TOPff}
{\cal A}_F({\bf B}_c \to {\bf B}_n M)&=&T({\bf B}_n)_i^n({\bf B}_c)_nH(\overline{15})^{li}_mM^m_l + C ({\bf B}_n)_{i}^n({\bf B}_c)_{n}H(\overline{15})^{il}_mM^m_l + {\cal A}_F({\cal O}(6)) \nonumber\\
&=&(T+C)({\bf B}_n)_i^n({\bf B}_c)_nH(\overline{15})^{il}_mM^m_l + {\cal A}_F({\cal O}(6))
\end{eqnarray}
where ${\cal A}_F({\cal O}(6))$ corresponds to the factorizable amplitude from ${\cal O}(6)$.
Here, we have used that $H(\overline{15})$ is symmetry in upper indices in the second line of Eq.~\eqref{TOPff}.
By comparing Eq.~(\ref{TOPff}) with Eq.~(\ref{Tamp}), we immediately identify that
only the  $a_6$ term in Eq.~(\ref{Tamp}) contains  the factorizable amplitude of ${\cal O}(\overline{15}).$\footnote{In general,  the term associated with
$a_6$ also contribute the non-factorizable part.} Consequently, we can safely neglect the  $a_4$, $a_5$ and $a_7$ terms
in $T_{{\cal O}_{\overline{15}}}$ of Eq.~(\ref{Tamp}) as they do not have the factorizable  contributions to the processes.
We remark  that in Eq.~(\ref{TOPff}), if $T$ and $C$ both exist  in ${\cal A}_F({\cal O}(\overline{15}))$,
one of them  should be canceled out by the corresponding term in ${\cal A}_F({\cal O}(6))$, resulting
in the process to be either color allowed or color suppressed. This can be explicitly demonstrated by the recent work in Ref.~\cite{He:2018joe} 
on the connection between the topological and $SU(3)_F$  approaches.

To illustrate the  effect of the  only $a_6$ term from ${\cal O}(\overline{15})$, we show the decay amplitudes of $\Lambda_c^+ \to p \pi^0$ and $\Lambda_c^+ \to n \pi^+$, given by~\cite{third}
 \begin{eqnarray}
 \label{LcTopn}
{\cal A}(\Lambda_c^+ \to p \pi^0) &\propto& \sqrt{2}\left(a_2+a_3-{a_6-a_7\over 2}\right)= \sqrt{2}\left(a_2+a_3-{a_6\over 2}\right)\,,
\nonumber\\
{\cal A}(\Lambda_c^+ \to n \pi^+) &\propto& 2\left(a_2+a_3+{a_6+a_7\over 2}\right)= 2\left(a_2+a_3+{a_6\over 2}\right)\,.
\end{eqnarray}
It is clear that the relation of ${\cal A}(\Lambda_c^+ \to n \pi^+) =\sqrt{2}{\cal A}(\Lambda_c^+ \to p \pi^0)$~\cite{Lu:2016ogy}
 is violated with the contributions from $a_6$. This violation has been explicitly pointed out in Ref.~\cite{Cheng:2018hwl} based on a dynamical model.
On the other hand, some direct relations still exist in some modes. For example,
one has that
\begin{eqnarray}
	{\cal A}(\Lambda_c^+ \to\Sigma^0 K^+) &\propto & \sqrt{2}\left(a_1 -a_3-\frac{a_4+a_5}{2}\right)=\sqrt{2}\left(a_1 -a_3 \right)\,,\nonumber\\
	{\cal A}(\Lambda_c^+\to \Sigma^+ K^0_S) &\propto& \sqrt{2}\left(a_1 -a_3-\frac{-a_4+a_5}{2}\right)=\sqrt{2}\left(a_1 -a_3\right)\,.
\end{eqnarray}
Future experimental searches for these decays will confirm if the  discussions based on $SU(3)_F$ are right or not.

We are now ready to perform our numerical calculation.
Since the $SU(3)_F$ flavor symmetry does not involve the  dynamical details,
we have to determine the parameters in the irreducible amplitude by the experimental data,
which can be found in the PDG~\cite{pdg} along with the recent measurements by BESIII~\cite{Ablikim:2017ors, Ablikim:2015flg}.
Currently, there are 9 data points from the absolute branching fractions, along with
the original data of
${\cal B}(\Lambda_c^+ \to p \pi^0)=(0.8\pm1.3)\times10^{-4}$ by BESIII~\cite{YKC},
which are summarized in Table~\ref{Table1}.
In addition, we  include the relative branching ratio of
${\cal R}_{\Xi_c^0}\equiv {\cal B}(\Xi_c^0 \to \Lambda^0 \bar{K}^0)/{\cal B}(\Xi_c^0 \to \Xi^-\pi^+)=0.420\pm0.056$ in our fitting.
\begin{table}
\caption{Decay amplitudes related to the $SU(3)_F$ parameters and the experimental data for the absolute branching fractions and
${\cal R}_{\Xi_c^0}$~\cite{pdg,Ablikim:2017ors,Ablikim:2015flg,YKC}.}
\begin{center}
{
\begin{tabular}[t]{ccc}
\hline
Channel &Amplitude& Data \\
\hline
$ \Lambda_{c}^{+}  \to  \Sigma^{+} \pi^{0} $ & $ \sqrt{2}(a_1 - a_2 - a_3) $ &$(12.4\pm1.0)\times 10^{-3}$\\
$ \Lambda_{c}^{+}  \to  \Sigma^{+} \eta $ & $ \frac{\sqrt{6}}{3}(-a_1 - a_2 + a_3) $ &$(7.0\pm2.3)\times 10^{-3}$\\
$ \Lambda_{c}^{+}  \to  \Sigma^{0} \pi^{+} $ & $ \sqrt{2}(-a_1 + a_2 + a_3) $ &$(12.9\pm0.7)\times 10^{-3}$\\
$ \Lambda_{c}^{+}  \to  \Xi^{0} K^{+} $ & $ -2a_2 $ &  $(5.9\pm1.0)\times 10^{-3}$\\
$ \Lambda_{c}^{+}  \to  p \bar{K}^{0} $ & $ -2a_1 + a_6 $ &$(31.6\pm1.6)\times 10^{-3}$\\
$ \Lambda_{c}^{+}  \to  \Lambda^{0} \pi^{+} $ & $ \frac{\sqrt{6}}{3}(-a_1 - a_2 - a_3 - a_6) $ &$(13.0\pm0.7)\times 10^{-3}$\\
\hline
$ \Lambda_{c}^{+}  \to  \Sigma^{0} K^{+} $ & $ \sqrt{2}(a_1 -a_3) $ &$(5.2\pm0.8)\times 10^{-4}$\\
$ \Lambda_{c}^{+}  \to  \Sigma^{+} K^{0} $ & $2(a_1 -a_3) $ &-\\
$ \Lambda_{c}^{+}  \to  p \pi^{0} $ & $ \sqrt{2}(a_2 + a_3 - \frac{a_6}{2}) $ &$(0.8\pm1.3)\times 10^{-4}$\\
$ \Lambda_{c}^{+}  \to  n \pi^{+} $ & $ 2(a_2 + a_3 + \frac{a_6}{2}) $ &-\\
$ \Lambda_{c}^{+}  \to  p \eta $ & $ \frac{\sqrt{6}}{3}(-2a_1 + a_2 - a_3 + \frac{3}{2}a_6) $ &$(12.4\pm3.0)\times 10^{-4}$\\
$ \Lambda_{c}^{+}  \to  \Lambda^{0} K^{+} $ & $ \frac{\sqrt{6}}{3}(a_1 - 2a_2 + a_3 + a_6) $ &$(6.1\pm1.2)\times 10^{-4}$\\
\hline
$ \Xi_{c}^{0}  \to  \Xi^{-} \pi^{+} $ & $ 2a_1 + a_6 $ &-\\
$ \Xi_{c}^{0}  \to  \Lambda^{0} \bar{K}^{0} $ & $ \frac{\sqrt{6}}{3}(-2a_1 + a_2 + a_3 + \frac{a_6}{2}) $ &-\\
${\cal R}_{\Xi_c^0}$ & & $0.420\pm0.056$\\
\hline
\label{Table1}
\end{tabular}
}
\end{center}
\end{table}
Altogether, there are seven $SU(3)_F$ parameters ($a_1,|a_2|e^{i\delta_{a_2}},|a_3|e^{i\delta_{a_3}},|a_6|e^{i\delta_{a_6}}$) to fit with
eleven data pointes  in Table~\ref{Table1}. Here, we have set $a_1$ to be a real parameter due to the removal of an overall phase.
We use the minimum $\chi^2$ fit as shown in Ref.~\cite{second}.
Explicitly, we  obtain
\begin{eqnarray}\label{ai}
  (a_1,|a_2|,|a_3|,|a_6|) &=& (0.271\pm0.006 , 0.126\pm0.010, 0.051\pm0.012, 0.055\pm0.030)GeV^3\,,
   \nonumber\\
  (\delta_{a_2},\delta_{a_3},\delta_{a_6}) &=& (82\pm6, -20 \pm 24, 40\pm36)^{\circ}\,,
    \nonumber\\
\chi^2/d.o.f &=&  0.5\,,
\end{eqnarray}
where $d.o.f$ represents the degree of freedom.
The value of $\chi^2/d.o.f$ indicates that our fit is good.
In the previous studies of the $\Lambda^+_c$ decays based on $SU(3)_F$~\cite{Lu:2016ogy,zero,first,second,third},
 the contributions of ${\cal O}(\overline{15})$ have been neglected. Our results in Eq.~\eqref{ai} show
 that the absolute value of $a_6$ is about $1/6$ compared to that of the leading one $a_1$ in  $T({\cal O}_6)$,
so that the ignorance of ${\cal O}(\overline{15})$ is indeed valid.

   In Table~\ref{Table2}, we list
   our fitting results for the branching ratios of the Cabibbo allowed and singly Cabibbo suppressed $\Lambda_{c}^+$ decays.
  In the table, we have also included the previous results based on $SU(3)_F$~\cite{second} without ${\cal O}(\overline{15})$ along with the data as well as
 those from the dynamical model calculations by CKX~\cite{Cheng:2018hwl}.
\begin{table}[t]
    \caption{Branching ratios for the Cabibbo allowed and singly Cabibbo suppressed decays of $\Lambda_{c}^+$.
}
\begin{center}
 \setlength{\tabcolsep}{2mm}{
\begin{tabular}{ccccc}
  \hline
    Decay branching ratio & This work & Data & $SU(3)_F$~\cite{second}&CKX~\cite{Cheng:2018hwl} \\
  \hline
  $10^3{\cal B}(\ \Lambda_{c}^{+}  \to  \Sigma^{+} \pi^{0}) $ &$ 12.6 \pm 2.1 $&$12.4\pm1.0$ & $12.8 \pm 2.3$  &-\\
$ 10^3{\cal B}(\Lambda_{c}^{+}  \to  \Sigma^{+} \eta) $ & $5.4 \pm 1.0 $&$7.0\pm2.3$ &$7.1 \pm 3.8 $&-\\
$ 10^3{\cal B}(\Lambda_{c}^{+}  \to  \Sigma^{0} \pi^{+}) $ & $ 12.6 \pm 2.1 $&$12.9\pm0.7$ & $12.8 \pm 2.3 $&-\\
$10^3{\cal B}( \Lambda_{c}^{+}  \to  \Xi^{0} K^{+}) $& $  5.9 \pm 1.0  $&$5.9\pm0.9$ & $5.5 \pm 1.4$&-\\
$ 10^3{\cal B}(\Lambda_{c}^{+}  \to  p \bar{K}^{0}) $ & $  31.3 \pm 1.6$&$31.6\pm1.6$ & $32.7 \pm 1.5$&-\\
$ 10^3{\cal B}(\Lambda_{c}^{+}  \to  \Lambda^{0} \pi^{+}) $  &$ 13.1 \pm 1.6 $&$13.0\pm0.7$ &$12.8\pm 1.7$ &-\\
\hline
  $10^4{\cal B}( \Lambda_{c}^{+}  \to  \Sigma^{+} K^{0}) $ & $ 11.4 \pm 2.0 $&- & $8.0 \pm 1.6$&14.4\\
$ 10^4{\cal B}(\Lambda_{c}^{+}  \to  \Sigma^{0} K^{+}) $ & $ 5.7 \pm 1.0 $ &$5.2\pm0.8$& $4.0 \pm 0.8$ &7.18\\
$10^4{\cal B}(\Lambda_c^+\to p \pi^0)$ & $1.3\pm0.7$ &$<2.7 $& $5.7\pm 1.5$ &$0.8$\\
$ 10^4{\cal B}(\Lambda_{c}^{+}  \to  p \eta $) & $ 13.0 \pm 1.0 $&  $12.4\pm3.0$ & $12.5 ^{+ 3.8 }_{- 3.6 }$& 12.8\\
$10^4{\cal B}(\Lambda_c^+ \to n \pi^+ )$&$ 6.1\pm2.0 $& -&$11.3\pm2.9 $&$2.7$\\
$10^4{\cal B}( \Lambda_{c}^{+}  \to  \Lambda^{0} K^{+}) $ & $ 6.4 \pm 0.9 $&$6.1\pm1.2$ & $4.6 \pm 0.9$&10.6\\
  \hline
\label{Table2}
\end{tabular}
}
\end{center}
\end{table}
As seen in Table~\ref{Table2}, our  results for
 the Cabibbo allowed   $\Lambda_{c}^+$ decays with the consideration of ${\cal O}(\overline{15})$
 are slightly better than those without ${\cal O}(\overline{15})$, but they all fit the data well.
 On the other hand,  the decay branching ratios for singly Cabibbo suppressed modes of  $\Lambda_{c}^+$
 with and without ${\cal O}(\overline{15})$ are quite different. In particular, we predict that
 ${\cal B}(\Lambda_c^+\to p \pi^0)=(1.3\pm0.7)\times 10^{-4}$, which is consistent with the experiments upper limit
 of   $2.7\times 10^{-4}$ as well as the result of $0.8\times 10^{-4}$ calculated by the pole model with current algebra in Ref.~\cite{Cheng:2018hwl}.
 It is clear that the  inconsistent branching
 ratio of  $(5.7\pm 1.5)\times 10^{-4}$ in the previous study with $SU(3)_F$~\cite{second}  results from
 the ignorance of ${\cal O}(\overline{15})$, in which
 a large destructive interference occurs between ${\cal O}(\overline{15})$ and ${\cal O}(6)$.
 It is also interesting to note that ${\cal B}(\Lambda_c^+ \to n \pi^+)$ is found  to be $(6.1\pm2.0)\times 10^{-4} $, which is reduced 
 by almost a factor 2 in comparing with that in Ref.~\cite{second}. Although the signs for the contributions  
 from $a_6$ to  $\Lambda_c^+\to p \pi^0$ and $\Lambda_c^+\to n \pi^+$  in Eq.~(\ref{LcTopn})  are opposite,
 the resulting values are both reduced due to the complex numbers of $a_{2,3}$ and $a_6$ in Eq.~(\ref{ai}).
 
In addition, from Table~\ref{Table2}, we have that
	\begin{equation}\label{bench}
	{\cal B}(\Lambda_c^+\to \Sigma^+ K^0_S)=(5.7\pm 1.0)\times 10^{-4}\,,
	\end{equation}
which agrees with  the experimental value of ${\cal B}(\Lambda_c^+ \to \Sigma^0 K^+)=(5.2\pm 0.8)\times 10^{-4}$~\cite{pdg}.
The future search for $\Lambda_c^+\to \Sigma^+ K^0_S$ is a good test for $SU(3)_F$.
 
 Finally, we remark that we are unable to discuss the $SU(3)_F$ breaking effects after including the contributions of  ${\cal O}(\overline{15})$
 in the fit due to the insufficient experimental data points. Once more experimental data
 are available  in the future, the studies of these effects along with the $\eta'$ channels would be possible.

In sum, we have studied the two-body decays of $\Lambda_c^+\to {\bf B}_nM$ based on the  approach with the $SU(3)_F$ flavor symmetry,
which is a powerful tool to examine charmed baryon physics and  allows us to connect the physical quantities without knowing the underlined dynamics.
We have successfully fitted all the existing experimental data from the  Cabibbo allowed and suppressed   decays of $\Lambda_{c}^+$.
By considering the approach with the topological quark diagrams,
for the first time, the contributions from ${\cal O}(\overline{15})$ have been included in the calculations with the $SU(3)_F$ method.
As a result, we have predicted all singly Cabibbo  suppressed  decays. In particular,
we have found that ${\cal B}(\Lambda_c^+\to p \pi^0)=(1.3\pm0.7)\times 10^{-4}$, which is slightly below the current experimental upper
limit of $2.7\times 10^{-4}$. This result can be tested by the  experiments at BESIII and Belle-II.

\section*{ACKNOWLEDGMENTS}
This work was supported in part by National Center for Theoretical Sciences and
MoST (MoST-104-2112-M-007-003-MY3 and MoST-107-2119-M-007-013-MY3).


\begin{thebibliography}{99}
\bibitem{Belle}
  S.~B.~Yang {\it et al.} [Belle Collaboration],
  Phys.\ Rev.\ Lett.\  {\bf 117},  011801 (2016)


\bibitem{Ablikim:2015flg}
M.~Ablikim {\it et al.} [BESIII Collaboration],
Phys.\ Rev.\ Lett.\  {\bf 116}, 052001 (2016).

\bibitem{Ablikim:2015prg}
  M.~Ablikim {\it et al.} [BESIII Collaboration],
  Phys.\ Rev.\ Lett.\  {\bf 115},  221805 (2015)

\bibitem{Ablikim:2016tze}
  M.~Ablikim {\it et al.} [BESIII Collaboration],
  Phys.\ Rev.\ Lett.\  {\bf 117},  232002 (2016);

\bibitem{Ablikim:2016mcr}
  M.~Ablikim {\it et al.} [BESIII Collaboration],
  Phys.\ Rev.\ Lett.\  {\bf 118}, 112001 (2017).

\bibitem{Ablikim:2017ors}
M.~Ablikim {\it et al.} [BESIII Collaboration],
Phys.\ Rev.\ D {\bf 95}, 111102 (2017).

\bibitem{Ablikim:2016vqd}
  M.~Ablikim {\it et al.} [BESIII Collaboration],
  Phys.\ Lett.\ B {\bf 767}, 42 (2017)

\bibitem{Ablikim:2017iqd}
  M.~Ablikim {\it et al.} [BESIII Collaboration],
  Phys.\ Lett.\ B {\bf 772}, 388 (2017)

\bibitem{Ablikim:2018jfs}
  M.~Ablikim {\it et al.} [BESIII Collaboration],
  Phys.\ Rev.\ Lett.\  {\bf 121},  062003 (2018).

  \bibitem{Ablikim:2018bir}
  M.~Ablikim {\it et al.} [BESIII Collaboration],
  Phys.\ Lett.\ B {\bf 783}, 200 (2018)

\bibitem{pdg}
M. Tanabashi
{\it et al.} [Particle Data Group],
Phys. Rev. D {\bf 98}, 030001 (2018).

\bibitem{Aaij:2015bfa}
R.~Aaij {\it et al.} [LHCb Collaboration],
Nature Phys.\  {\bf 11}, 743 (2015). 

\bibitem{Hsiao:2018zqd}
  Y.~K.~Hsiao and C.~Q.~Geng,
  Phys.\ Lett.\ B {\bf 782}, 728 (2018).


\bibitem{Cheng:2015iom}
  H.~Y.~Cheng,
  ``Charmed baryons circa 2015,''
  Front.\ Phys.\ (Beijing) {\bf 10},  101406 (2015).

 \bibitem{Uppal:1994pt}
  T.~Uppal, R.~C.~Verma and M.~P.~Khanna,
  Phys.\ Rev.\ D {\bf 49}, 3417 (1994).


 \bibitem{Chen:2002jr}
  S.~L.~Chen, X.~H.~Guo, X.~Q.~Li and G.~L.~Wang,
  Commun.\ Theor.\ Phys.\  {\bf 40}, 563 (2003).

  \bibitem{Cheng:2018hwl}
  H.~Y.~Cheng, X.~W.~Kang and F.~Xu,
  Phys.\ Rev.\ D {\bf 97}, no. 7, 074028 (2018).

   \bibitem{Sharma:1996sc}
  K.~K.~Sharma and R.~C.~Verma,
  Phys.\ Rev.\ D {\bf 55}, 7067 (1997).

 \bibitem{Lu:2016ogy}
C.D.~Lu, W.~Wang and F.S.~Yu,
Phys.\ Rev.\ D {\bf 93}, 056008 (2016).

 \bibitem{first}
  C.~Q.~Geng, Y.~K.~Hsiao, C.~W.~Liu and T.~H.~Tsai,
  JHEP {\bf 1711}, 147 (2017).

 \bibitem{zero}
  C.~Q.~Geng, Y.~K.~Hsiao, Y.~H.~Lin and L.~L.~Liu,
  Phys.\ Lett.\ B {\bf 776}, 265 (2018).

\bibitem{second}
  C.~Q.~Geng, Y.~K.~Hsiao, C.~W.~Liu and T.~H.~Tsai,
  Phys.\ Rev.\ D {\bf 97}, no. 7, 073006 (2018).
\bibitem{third}
  C.~Q.~Geng, Y.~K.~Hsiao, C.~W.~Liu and T.~H.~Tsai,
  Eur.\ Phys.\ J.\ C {\bf 78}, no. 7, 593 (2018).

\bibitem{Savage:1989qr}
M.J.~Savage and R.P.~Springer,
Phys.\ Rev.\ D {\bf 42}, 1527 (1990).

\bibitem{Savage:1991wu}
M.J.~Savage,
Phys.\ Lett.\ B {\bf 257}, 414 (1991).

\bibitem{fourth}
  C.~Q.~Geng, Y.~K.~Hsiao, C.~W.~Liu and T.~H.~Tsai,
  arXiv:1810.01079 [hep-ph].

\bibitem{old1}
M.~K.~Gaillard and B.~W.~Lee, Phys.\ Rev.\ Lett.\  {\bf 33}, 108 (1974).

\bibitem{old2}
G..~Altaerelli  and L.~Maiani, Phys.\ Lett.\  {\bf B 52}, 351 (1974).


\bibitem{Abbott:1979fw}
  L.~F.~Abbott, P.~Sikivie and M.~B.~Wise,
  Phys.\ Rev.\ D {\bf 21} (1980) 768.

\bibitem{Zeppenfeld:1980ex}
  D.~Zeppenfeld,
  Z.\ Phys.\ C {\bf 8}, 77 (1981).


\bibitem{deSwart:1963pdg}
  J.~J.~de Swart,
  Rev.\ Mod.\ Phys.\  {\bf 35}, 916 (1963)
  Erratum: [Rev.\ Mod.\ Phys.\  {\bf 37}, 326 (1965)].
  doi:10.1103/RevModPhys.35.916


  \bibitem{Chau:1995gk}
  L.~L.~Chau, H.~Y.~Cheng and B.~Tseng,
  Phys.\ Rev.\ D {\bf 54}, 2132 (1996)

\bibitem{He:2018joe}
  X.~G.~He, Y.~J.~Shi and W.~Wang,
  arXiv:1811.03480 [hep-ph].



\bibitem{Buras:1998raa} 
A.J.~Buras, hep-ph/9806471.

\bibitem{Cheng:1991sn}
H.Y.~Cheng and B.~Tseng,
Phys.\ Rev.\ D {\bf 46}, 1042 (1992); {\bf 55}, 1697(E) (1997).

\bibitem{YKC}
Private communication with
the BESIII Collaboration.




\end{thebibliography}
\end{document}